\documentclass[11pt,onecolumn,draftcls]{IEEEtran}
\usepackage{graphicx}
\begin{document}

\title{Physical-Layer Security: Combining Error Control Coding and Cryptography
\footnote{
Willie K Harrison and Steven W. McLaughlin are with the School of ECE, Georgia Institute of Technology, Atlanta, GA.\\
\indent This work has been submitted and accepted to IEEE International Conference on Communications, and will be presented there June 14-18, 2009.}
}
\author{\authorblockN{Willie K Harrison and Steven W. McLaughlin}}

\maketitle

\begin{abstract}

In this paper we consider tandem error control coding and cryptography in the setting of the {\em wiretap channel} due to Wyner.  In a typical communications system a cryptographic application is run at a layer above the physical layer and assumes the channel is error free.  However, in any real application the channels for friendly users and passive eavesdroppers are not error free and Wyner's wiretap model addresses this scenario. Using this model, we show the security of a common cryptographic primitive, i.e. a keystream generator based on linear feedback shift registers (LFSR), can be strengthened by exploiting properties of the physical layer. A passive eavesdropper can be made to experience greater difficulty in cracking an LFSR-based cryptographic system insomuch that the computational complexity of discovering the secret key increases by orders of magnitude, or is altogether infeasible. This result is shown for two fast correlation attacks originally presented by Meier and Staffelbach, in the context of channel errors due to the wiretap channel model.

\end{abstract}

\section{Introduction}
\begin{figure*}
  \begin{center}
  \includegraphics[width=7in]{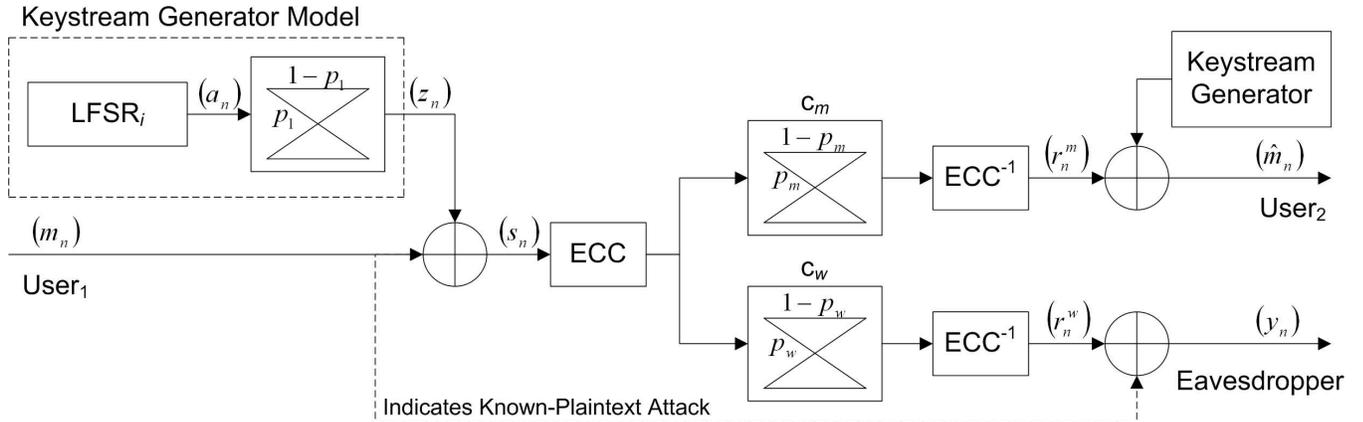}\\
  \end{center}
  \caption{Portrayal of a known-plaintext attack on the wiretap channel model where two friendly parties share information over a main channel $c_m$ and an eavesdropper observes communications through a wiretap channel $c_w$. In practice the keystream generator is comprised of $M$ LFSR output sequences combined according to a function $f$. It is simplified from its true condition and modeled as a single LFSR with a BSC.}\label{fig:bigSystem}
\end{figure*}

Traditionally communication systems have implemented security measures by cryptographic means. However, with the introduction of the wiretap channel model by Wyner \cite{Wyner75}, it became clear that security can also be achieved through means of channel coding. The wiretap channel model portrays two friendly users sharing information over a \emph{main} communications channel $c_m$ (e.g. a fading wireless channel \cite{Barros06}) and a passive eavesdropper observing a degraded version of the information through a \emph{wiretap} channel $c_w$. As in \cite{Wyner75}, we will assume that both channels are discrete and memoryless. Fig. \ref{fig:bigSystem} portrays this scenario using binary symmetric channel (BSC) models for both $c_m$ and $c_w$. If the communication over $c_m$ is of a private nature, it then becomes necessary to accomplish two seemingly conflicting tasks of reliability between the friendly users and security against the eavesdropper through some encoding technique. The purpose of this paper is to quantify the additional complexity that the eavesdropper faces when the security problem is addressed with channel errors at the physical layer in mind.

The existence of codes providing reliability to friendly parties while maintaining some level of confidentiality is crucial to increasing necessary computations for an eavesdropper, and has been proven by Wyner in \cite{Wyner75} as well as Csiszar and Corner in \cite{Csiszar78}. Practical codes of this kind, however, were not discovered until much later \cite{Wei91}. It has since been shown for many varying circumstances and channels that practical codes exist which satisfy both design constraints of reliability and secrecy. For example, it has been shown in \cite{McLaughlin07} that practical low-density parity-check (LDPC) codes exist which achieve these two criteria for a noiseless channel $c_m$ and a binary erasure channel $c_w$. Similar results have been shown in \cite{Bloch06}, also making use of LDPC codes as well as multilevel coding for the case of independent quasi-static fading channels $c_m$ and $c_w$. In this paper we address a practical scenario where both $c_m$ and $c_w$ are treated as BSCs with probabilities of a bit flip $p_m$ and $p_w$, respectively. It is assumed that the wiretap channel quality is less than that of the main channel, that is $p_w>p_m$. This might be the case, for example, in a zoned-security application where the friendly parties are inside a building and the eavesdropper is outside the building monitoring communications.

The rest of the paper is outlined as follows. First we give some discussion on the general setting.  We focus our attention on linear feedback shift register (LFSR) cryptographic applications because attacks against them have been well documented and we are able to quantify the increase in complexity that the eavesdropper experiences due to errors in the wiretap channel.  Two well-known attacks originally given in \cite{Meier89} will be considered, and it will be shown that an eavesdropper can be made to fail in obtaining the secret key in an otherwise successful scenario by considering the effects of channel errors presented by some physical means. The background for the LFSR-based cryptography is given in section \ref{sec:background}, while the attacks are presented briefly in section \ref{sec:algorithms}. Afterwards, section \ref{sec:proofOfConcept} provides evidence of a physical-layer of security under the conditions of the attacks presented in the previous section. Theoretical results as well as simulation output for the two attacks are also included in this section. Finally conclusions of these findings are provided in section \ref{sec:conclusion}.

\section{Background}\label{sec:background}
It has been shown by Shannon and others that a one-time pad can achieve \emph{perfect secrecy} as a cryptographic encoding technique \cite{Shannon49}, meaning that knowing the codeword or encoded sequence gives no information on the value of the original message. However, implementation of a one-time pad relies on a perfectly random key sequence. Assuming that a user is capable of generating this sequence of elements, the problem of communicating with absolute secrecy can be solved, but at the expense of requiring distribution of a secret key which is the same length as the original message \cite{Welsh}.

Due to the issue of key distribution inherent in the one-time pad encoding mechanism, other methods are used to attempt to emulate the secrecy aspects of the one-time pad while providing a more practical key length. One such system is given in \cite{Meier89}, \cite{Siegenthaler85}, \cite{ChepyzhovS91}, and \cite{JohanssonJ02}. The encoder for this system is comprised of a keystream generator that produces a pseudorandom key sequence $(z_n)$ by combining $M$ LFSR output sequences using a function $f$. The notation \mbox{$(z_n)=(z_0,z_1,\ldots)$} denotes a sequence or vector whose $n$th element is $z_n$. Assuming all data sequences to be binary, a ciphertext bit sequence $(s_n)$ is produced using a bit-wise exclusive or (XOR) operation between the message sequence $(m_n)$ and the keystream sequence $(z_n)$, as portrayed in Fig. \ref{fig:bigSystem}. The sequence $(a_n)$ is the output sequence of a single LFSR, say the $i$th one. The effective \emph{key} of the system consists of the initial conditions of the $M$ shift registers, and hence is fixed in length regardless of the length of $(m_n)$. Decoding is accomplished using the XOR operation with the same keystream sequence $(z_n)$, which friendly parties can duplicate once they know the key. If it is assumed that the bits of $(z_n)$ are random independent and identically distributed (i.i.d.), and therefore that bits in the sequence cannot be predicted by an eavesdropper, then the system achieves the secrecy of the one-time pad.

This assumption is untrue, however, in many instances. For example, Siegenthaler showed using only ciphertext that the secret key (initial state) of a contributing LFSR can be obtained by calculating a correlation metric for all possible initial conditions of the LFSR, and then comparing to a Neyman-Pearson threshold determined by the statistics of the data \cite{Siegenthaler85}. While this particular attack requires $2^k-1$ correlation calculations, fast-correlation techniques exist where it is shown that a low-weight connection polynomial of an LFSR, i.e. one with a small number of feedback loops, produces a more susceptible system to correlation methods \cite{Meier89, ChepyzhovS91}. Despite the shortcomings of LFSR-based generators, they continue to be used in modern cryptographic systems, including $E_0$ the system employed by Bluetooth \cite{Bluetooth07}. This is the case due to the relative ease in computations that an LFSR-based system provides. Many wireless and handheld technologies benefit from LFSR-based cryptography.

\begin{figure}
  \begin{center}
  \includegraphics[width=3.34in]{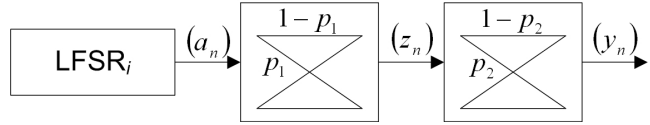}\\
  \end{center}
  \caption{Wiretap channel model flow diagram relating sequences $(a_n)$, $(z_n)$, and $(y_n)$ using a pair of binary symmetric channels.}\label{fig:model2BSCs}
\end{figure}

The attack of the LFSR-based cryptographic primitive assumes that the keystream sequence $(z_n)$ is correlated to the output sequence of the $i$th LFSR $(a_n)$ with correlation value $1-p_1$, and thus can be modeled as a BSC with \mbox{$\Pr{(a_j\neq z_j)}=p_1$} for $j=0,1,\ldots,N-1$, where $N$ is taken to be the length of an observed sequence. Fig. \ref{fig:bigSystem} shows this modeling of the keystream generator. A known-plaintext attack is portrayed in the figure where an eavesdropper has some means of obtaining $N$ bits of the original message; therefore, if the sequence $(s_n)$ is observed without error, then the first $N$ bits of the keystream sequence $(z_n)$ can be reconstructed exactly. It is assumed that $p_w>p_m$ implying more errors in the wiretap channel than in the main channel; therefore, an encoding technique is chosen to guarantee reliable communications between friendly parties while maintaining some percentage of bit errors in the wiretap channel. The effective error rate after applying error control coding (ECC) in the wiretap channel is given as $p_2$, and the model considered for the eavesdropper is simplified to that shown in Fig. \ref{fig:model2BSCs}. This figure indicates a pair of BSCs, where the first models the correlation of the sequences $(a_n)$ and $(z_n)$, and the second models bit errors in the wiretap channel after channel decoding. The output sequence of the final BSC $(y_n)$ is obtained in practice by applying the $N$ known bits of $(m_n)$ to the decoded sequence as shown in Fig. \ref{fig:bigSystem}. This sequence can be thought of as a noisy version of $(a_n)$, with a single BSC separating the two sequences. The probability of a bit flip in this BSC is denoted $p'$ and is calculated to be
\begin{equation}\label{eq:pPrime}
    p'=p_1(1-p_2)+(1-p_1)p_2=p_1+p_2-2p_1p_2.
\end{equation}
The cryptographic system is said to be compromised if an eavesdropper can obtain the initial contents of the $i$th LFSR using $(y_n)$ assuming knowledge of the LFSR connection polynomial is public.

\section{Cryptographic Algorithms}\label{sec:algorithms}
Both attacks presented in \cite{Meier89} reconstruct the key of the $i$th LFSR using checks which are derived from the  feedback polynomial $g(x)$. This polynomial governs the structure of the LFSR, and guarantees a maximal-length output sequence before repeating if and only if $g(x)=g_0+g_1x+g_2x^2+\cdots +g_kx^k$ is primitive in GF(2), where $g_j \in \{0,1\}$ for $j=0,1,\ldots,k$ \cite{MoonArches}. Define $t$ to be the number of feedback loops in the LFSR. For primitive $g(x)$ of order $k$, $g_0=g_k=1$ and the total number of nonzero coefficients of $g(x)$ is odd \cite{Meier89}, thus providing an even value of $t$ ($g_k$ does not \emph{feed back}). Let the indices of the nonzero coefficients in $g(x)$ be denoted $j_0,j_1,\ldots,j_t$; then $j_0=0$ and $j_t=k$. Now consider the $j$th bit of the sequence $(a_n)$. Due to the structure of $g(x)$, $a_{j+j_0} + a_{j+j_1} + \cdots + a_{j+j_t}=0$. This expression is calculated in GF(2), and thus simplifies to
\begin{equation}\label{eq:check}
    a_{j} = a_{j+j_1} + a_{j+j_2} + \cdots + a_{j+j_t}.
\end{equation}
Except for those within $t$ bits of the end of the sequence, every bit can be expected to contribute to $t+1$ checks of this kind. Additional checks are generated using a rule sometimes referred to as freshman exponentiation which states that for elements $x$ and $y$ in GF(2), $(x+y)^2=x^2+y^2$ \cite{MoonArches}. Check expressions given by (\ref{eq:check}) can then be repeatedly squared until limited by the length of the sequence $N$, providing additional check expressions with each squaring. Both attacks rely on computing these checks using the bits of $(y_n)$, and counting checks which hold with equality. Of course a check can still hold if an even number of bits in a check expression have been flipped, hence bits are assigned conditional probabilities of being correct given the number of satisfied checks. These probabilities are stored in the vector $(p^*_n)$. Let the number of satisfied checks containing $y_j$ be denoted as $c_s^j$, while the number of total checks for which $y_j$ plays a role is expressed as $c_{to}^j$. If $c_s^j=h$ and $c_{to}^j=m$, then
\begin{equation}\label{eq:pstar}
 \begin{array}{ll}
  p^*_j & =\Pr{(y_j=a_j|c_s^j=h,c_{to}^j=m)}\\
   & =\frac{p's^h(1-s)^{m-h}}{p's^h(1-s)^{m-h}+(1-p')(1-s)^hs^{m-h}},
 \end{array}
\end{equation}
where $s$ is defined as the probability that an even number of errors occur in the bits of the check expression discounting $y_j$ \cite{Meier89}. This value can be calculated recursively as \mbox{$s(j) = (1-p')s(j-1) + p'(1-s(j-1))$} where \mbox{$s(1) = 1-p'$} and \mbox{$s=s(t)$}.

\subsection{Attack A}\label{subsec:algA}
The first attack in \cite{Meier89} is founded on the principle that bits which satisfy the most checks are the most reliable. Using the $k$ bits which have the greatest values in $(p^*_n)$, a system of equations is determined and solved where the solution is the key or initial contents of the LFSR. This system of equations is constructed using the fact that every output of an LFSR is merely a linear combination of the bits in the initial state. The key is obtained by solving the system using a method such as LU decomposition tailored to operations in GF(2) \cite{MoonBlack}. Measures must be taken to ensure that the group of $k$ bits chosen have linearly independent key bit combinations.

In order to determine whether the obtained solution is the key, a threshold for a correlation metric between $(y_n)$ and a sequence generated by the solution to the system of equations must be formed \cite{Siegenthaler85}. If the solution is determined to be incorrect by the threshold comparison, the algorithm must then perform an exhaustive search on possible error combinations in the $k$ chosen bits. The calculations necessary to perform this task dominate the performance of the algorithm, and hence define the computational complexity of attack A. Variations of the $k$ bits with Hamming distance 1,2,\ldots,$k$ are tried until a key is found which satisfies the correlation condition. In order to calculate a worst-case scenario, it is assumed that the eavesdropper is always able to detect a correct key.

\subsection{Attack B}\label{subsec:algB}
The second attack presented in \cite{Meier89} also makes use of the conditional probabilities $(p^*_n)$; however, the iterative nature of this attack alters these calculations slightly. Attack B is extremely comparable to Gallager's LDPC decoding algorithm \cite{Gallager63}. In the attack all conditional probabilities in the sequence $(p^*_n)$ are calculated using (\ref{eq:pstar}). A threshold $p_{thr}$ is derived by calculating the best possible increase in correct bits assuming that all bits with probability less than the threshold are flipped. This correction threshold is set to the value where any bit $y_j$ with $p^*_j<p_{thr}$ has a maximum likelihood of being incorrect. If a certain predetermined number of bits $N_{thr}$ have values in $(p^*_n)$ less than $p_{thr}$, then those bits are flipped. Otherwise the conditional probabilities $p^*_j$ for $j=0,1,\ldots,N-1$ are recalculated by exchanging the a priori probability $p'$ with the previous value of $p^*_j$ in (\ref{eq:pstar}). After a few iterations of probabilities, or once at least $N_{thr}$ untrustworthy bits are found, the bits are flipped and the algorithm continues in this way until a solution is obtained.

\section{Proof of Concept with Simulation Results}\label{sec:proofOfConcept}
If a channel encoding technique can guarantee bit errors for a passive eavesdropper regardless of ECC, then these errors can clearly contribute to the overall security of the system. The questions then remain of how to quantify the amount of security gained, and what value of $p_2$ will prevent an eavesdropper from gaining advantage in a correlation-based attack. To provide answers to these two questions, metrics used in \cite{Meier89} are analyzed. First in the case of attack A, suppose there are exactly $r$ errors in the $k$ chosen bits. Then the maximum number of iterations in an exhaustive search is
\begin{equation}\label{eq:maxIts}
    A(k,r)=\sum_{i=0}^r{k \choose i}\leq 2^{H(r/k)k}.
\end{equation}
The inequality makes use of the binary entropy function $H(x)$, and is well known. Of course $r$ is not readily available in practice, but it can be estimated making use of (\ref{eq:pstar}) in the expression $\bar{r}=k(1-\Pr{(y_j=a_j|c_s^j=h',c_{to}^j=m')})$, where $m'$ is the average number of checks relevant to any one bit, and $h'$ is the maximum integer such that $k$ bits exist which are expected to satisfy at least $h'$ checks. Therefore, given that the best $k$ bits are chosen, $\bar{r}$ of them are still expected to be in error. An estimate on an upper bound of the number of trials required is then given as $2^{H(\bar{r}/k)k}$.

\begin{figure}
  \begin{center}
  \includegraphics[width=90 mm]{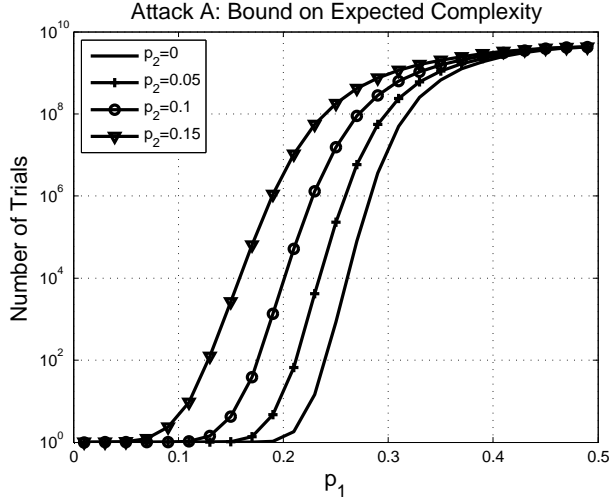}\\
  \end{center}
  \caption{Expected bound on the number of trials required to find the secret key using attack A for $k=32$, $N=k\times 10^6$, and $t=6$.}\label{fig:attackAtheoryBig}
\end{figure}

\begin{figure}
  \begin{center}
  \includegraphics[width=90 mm]{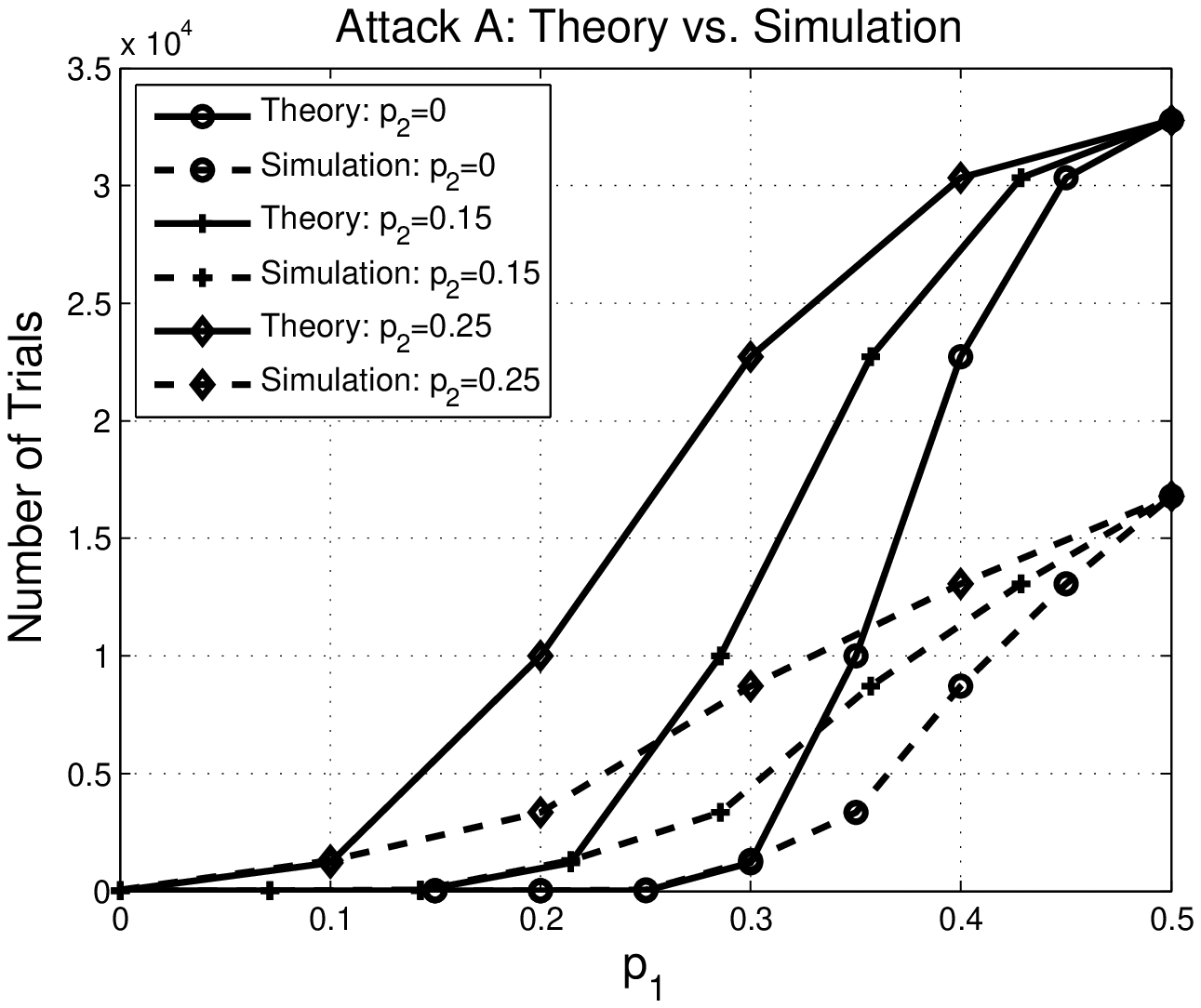}\\
  \end{center}
  \caption{Results from simulations of attack A showing necessary computations to crack the LFSR-based cryptographic system. Here $k=15$, $N=k\times 100$, and $t=4$.}\label{fig:simulationA}
\end{figure}

Fig. \ref{fig:attackAtheoryBig} shows this bound for varying $p_2$ values in an example featuring a length-32 LFSR while assuming that $N = 32\times10^6$ bits of $(m_n)$ are known by the eavesdropper. The greater the length of the observed data sequence, the easier the system will yield to a correlation-based attack. Simulations of the attack are compared with this bound and can be seen in Fig. \ref{fig:simulationA}, although for a shorter LFSR. Both the theoretical bounds and the simulations show channel conditions where attacks are expected to require a significant amount of additional computations due to nonzero $p_2$. As shown in Fig. \ref{fig:simulationA}, the expected bound is much tighter for smaller $p'$ values. Clearly as $p'$ approaches 0.5, attack A reverts to a brute-force attack which is expected to require $2^{k-1}$ iterations, while the bound approaches $2^k$. For smaller $p'$ the difference between the bound and the simulation results is not as pronounced. Regardless of this difference, when $k$ is large and $p'$ is close to 0.5 the task of finding the secret key becomes overwhelmingly expensive, and not feasible in many cases. Physical-layer considerations can be addressed in the choice of channel codes which can then drive $p'$ to 0.5 by increasing $p_2$, and thus obtain this extra level of security.

A similar analysis can be conducted for attack B; however, the attack has an underlying bipartite graph which connects check nodes to probabilities in $(p^*_n)$. Since the graph contains many cycles, after a few iterations probabilities become difficult to track; thus numbers of computations are likewise difficult to estimate. The strength of this attack is instead calculated by determining the effect of the first iteration of the algorithm. Recall that a threshold $p_{thr}$ was determined to maximize the probability that $y_j\neq a_j$ given that $p^*_j<p_{thr}$. Let $N_w$ be the expected number of bits such that both $a_j\neq y_j$ and $p^*_j<p_{thr}$, and let $N_v$ be the expected number of bits such that $a_j = y_j$ and $p^*_j<p_{thr}$, for $j=0,1,\ldots,N-1$. Also let $N_i=N_w-N_v$. If $N_{c0}$ represents the total number of bits such that $a_j=y_j$ prior to iteration, then the toggling of all bits with $p^*_j<p_{thr}$ will result in an expected $N_{c0}+N_i$ correct bits. Obviously if $N_i$ is negative, then the expected outcome of the first iteration will leave more bits in error than were originally so.

\begin{figure}
  \begin{center}
  \includegraphics[width=90 mm]{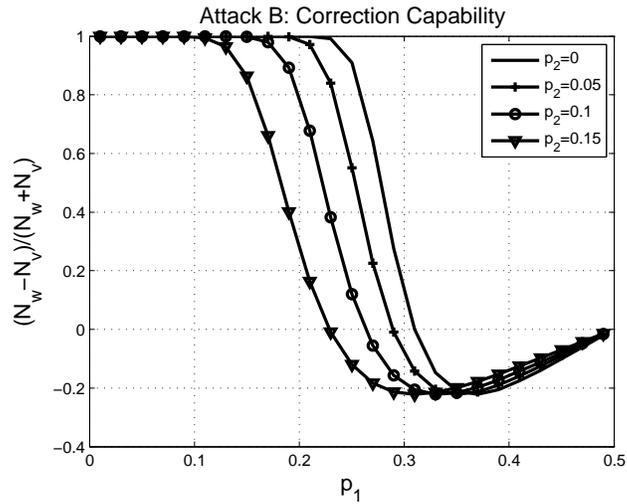}\\
  \end{center}
  \caption{Correction capability of attack B for $k=32$, $N=k\times 10^6$, and $t=6$. Negative values give a correction capability of zero indicating an inability to converge on the correct sequence.}\label{fig:attackBtheoryBig}
\end{figure}

The only way to ensure that the algorithm does not eventually converge on the correct sequence is to insist that attack B have no correction capability. While this would be impossible to guarantee under every scenario, we say that attack B has correction capability zero if $N_i<0$, i.e. $N_v>N_w$. The ratio
\begin{equation}\label{eq:correctionFactor}
    C=\frac{N_i}{N_w+N_v}
\end{equation}
is used to scale the value of $N_i$ to a real number in the range $[-1,1]$ while maintaining its sign. Fig. \ref{fig:attackBtheoryBig} shows the value of the correction ratio $C$ for several BSC parameters $p_2$, over a range of $p_1$ values. It should be noted that while a negative value of $C$ implies a correction capability of zero, conditions yielding positive $C$ values still may not converge on the correct sequence. Simulations of attack B have been consistent in a lack of convergence for cases where $C\leq 0$.

An example is in order. Let the primitive connection polynomial for the $i$th LFSR be written as $g(x)=x^{31}+x^{21}+x^{12}+x^3+x^2+x+1$ \cite{PolyWebsite}, and the correlation between $(a_n)$ and $(z_n)$ be 0.8, implying $p_1=0.2$. In the first of two cases $p_2=0$, meaning the eavesdropper is able to decode all channel errors in the wiretap channel using ECC, thus $p'=p_1=0.2$ and the correction ratio $C$ is calculated using (\ref{eq:correctionFactor}) to be 0.826. Case two assumes that $p_2=0.1$ indicating an error rate of 10 percent in $(y_n)$ which yields $p'=0.26$ by (\ref{eq:pPrime}), and $C=-0.034$ by (\ref{eq:correctionFactor}). Due to these values of $C$, it is expected that attack B will succeed in case one and fail in case two. Comparisons of the attacks are shown in Tab. \ref{tab:attackB}, where it is seen that case one converges on the correct output sequence in 16 rounds. Case two, however, requires 34 rounds before the algorithm stagnates and fails, a majority of rounds resulting in more bits in error than the previous round. Clearly an eavesdropper has been made to fail in an otherwise successful scenario due to the increased security inherent in the system which can be produced by wise implementation of channel coding.

\begin{table}
\caption{Simulation results of attack B comparing scenarios with and without added security due to the physical layer. For these simulations, $k=31$, $N=k\times 100$, $t=6$, and $p_1=0.2$.}
\begin{center}
\begin{tabular}{|c|c|c|c|c|}
  \hline

   & \multicolumn{2}{c|}{Case 1: $p_2=0$} & \multicolumn{2}{c|}{Case 2: $p_2=0.1$} \\ \hline
   & Number of  & Total & Number of & Total \\
   Round & bits flipped & correct bits & bits flipped & correct bits\\ \hline
  1 & 30 & 2487 & 1 & 2276 \\
  2 & 91 & 2526 & 3 & 2277 \\
  3 & 122 & 2586 & 6 & 2277 \\
  4 & 42 & 2628 & 8 & 2275 \\
  5 & 50 & 2676 & 11 & 2268 \\
  \vdots & \vdots & \vdots & \vdots & \vdots \\
  14 & 43 & 3075 & 2 & 2204 \\
  15 & 23 & 3098 & 100 & 2164 \\
  16 & 2 & 3100 & 4 & 2164 \\
  \vdots & - & - & \vdots & \vdots \\
  34 & - & - & 1 & 2079 \\
  35 & - & - & 0 & 2079 \\
  \vdots & - & - & 0 & 2079 \\
  \hline
\end{tabular}
\end{center}\label{tab:attackB}
\end{table}

\section{Conclusion}\label{sec:conclusion}
In conclusion, the wiretap channel model has been used to show security enhancements for wireless applications by considering the channel coding problem and the cryptography problem in tandem. These enhancements occur due to effects in the physical layer of a communications system. For a variety of applications where an eavesdropper experiences worse channel conditions than those between friendly parties, proper implementation of channel coding can ensure an increased difficulty in cracking cryptographic systems by preserving bit errors in the wiretap channel due to the physical layer. This principle was shown using an LFSR-based cryptographic system which is susceptible to correlation attacks in some cases. It has been shown using theory and simulations for two different attacks that channel coding can be used to either increase the difficulty of the attack or make it altogether impossible, thus providing a physical layer of security to the system.

\bibliographystyle{unsrt}
\bibliography{LFSRreferences}

\end{document}